%Paper: hep-th/9403070
%From: bdolan@thphys.may.ie
%Date: Fri, 11 Mar 94 18:19:37 GMT

\magnification=1200
\overfullrule= 0pt
\def\setup{\count90=0 \count80=0 \count91=0 \count85=0
\countdef\refno=80 \countdef\secno=85 \countdef\equref=\count90}
\def\R{\vrule height5.85pt depth.2pt \kern-.05pt \tt R}

\def\Box{\vbox{\hrule
                    \hbox{\vrule height 6pt \kern 6pt \vrule height 6pt}
                    \hrule}\kern 2pt}
\def\lo{\raise2pt\hbox{$<$}\kern-7pt\raise-2pt\hbox{$\sim$}}
\def\go{\raise2pt\hbox{$>$}\kern-7pt\raise-2pt\hbox{$\sim$}}
\def\parallel#1#2{\hbox{\kern1pt \vrule height#1pt
                                 \kern#2pt
                                 \vrule height#1pt \kern1pt}}

\def\vline{{\vrule height8pt depth4pt}\; }
\def\Vline{{\vrule height13pt depth8pt}\; }
\def\sub#1{{\lower 8pt \hbox{$#1$}}}

\def\Del{{\raise.5ex\hbox{$\bigtriangledown$}}}
\def\DEL#1{{\raise.5ex\hbox{$\bigtriangledown$}\raise 8pt \hbox{\kern -10pt
                 \hbox{$#1$}} }}
\def\autoeq{ {\global\advance\count90 by1} \eqno(\the\count90) }
\def\autoeql{ {\global\advance\count90 by1} & (\the\count90) }
\def\ceist{ {\global\advance\count91 by1} (\the\count91) }
\def\autosec{ {\global\advance\secno by 1} (\the\secno) }
\def\e{\hbox{e}}
\def\Lie#1{{\cal L}{\kern -6pt
            \hbox{\raise 1pt\hbox{-}}\kern 1pt} _{\vec{#1}}}

\def\Z{Z \kern-5pt \hbox{\raise 1pt\hbox{-}}\kern 1pt}
% THE FOLLOWING MACROS ARE FOR AUTOMATICALLY ARRANGING REFERENCES
% TO GET A REFERENCE USE \refread{#}
% IT REQUIRES A SINGLE ARGUMENT, WHICH IS THE REFERENCE # IN refs.tex
% ONCE \refread HAS BEEN INVOKED THE TITLE RESIDES IN \title
% THE AUTHOR IN \author AND THE DETAILS OF YEAR, JOURNAL OR
% PUBLISHER, VOLUME AND PAGE # IN \pub
\def\autoref{ {\global\advance\refno by 1} \kern -5pt [\the\refno]\kern 2pt}
\def\readref#1{{\count100=0 \openin1=refs\loop\ifnum\count100
<#1 \advance\count100 by1 \global\read1 to \title \global\read1 to \author
\global\read1 to \pub \repeat\closein1  }}
\def\reftitle{{ \kern -3pt \vtop{ \hbox{\title} \hbox{\author\ \pub} } }}
\def\ref{{ \kern -3pt\author\ \pub \kern -3.5pt }}

\def\refanon{{ \hbox{\pub}\kern -3.5pt }}

\def\circum#1{{ \kern -3.5pt $\hat{\hbox{#1}}$ \kern -3.5pt}}
\setup
\centerline{  }
\line{\hfill \hbox{DIAS-STP-94-05}}
\line{\hfill \hbox{February 1994} }
\vskip 2cm
\centerline{{\bf Co-variant Derivatives And The Renormalisation Group
Equation}
\footnote*{Work partly supported by an  Alexander von Humboldt
research stipendium.}}
\vskip 1.2cm
\centerline{Brian P. Dolan}
\vskip .5cm
\centerline{\it Department of Mathematical Physics, St. Patrick's College}
\centerline{\it Maynooth, Ireland}
\centerline{and}
\centerline{\it Dublin Institute for Advanced
Studies}
\centerline{\it 10, Burlington Rd., Dublin, Ireland}
\vskip .5cm
\centerline{e-mail:bdolan@thphys.may.ie}
\vskip 1.5cm
\centerline{ABSTRACT}
\noindent The renormalisation group equation for $N$-point correlation
functions can be interpreted in a geometrical manner as an equation
for Lie transport of amplitudes in the space of couplings.
The vector field generating the diffeomorphism has components
given by the $\beta$-functions of the theory. It is
argued that this simple picture requires modification whenever
any one of the points at which the amplitude is evaluated becomes
close to any other. This modification requires the introduction
of a connection on the space of couplings and new terms appear
in the renormalisation group equation involving co-variant derivatives
of the $\beta$-function and the curvature associated with the connection.
It is shown how the connection is related to the operator
expansion co-efficients, but there remains an arbitrariness
in its definition.

\vskip 1cm

%\vskip .5cm \noindent
%PACS Nos. $03.70.+$k and $11.10.G$h
\vfill\eject
{\bf \S 1 Introduction}
\vskip.5cm
Geometry has always played a central role in the development of
theoretical physics.
Recently a new possibility for the application of geometry to
physics has emerged, that is the use of geometrical concepts to
understand the \lq\lq space of theories",
\autoref\newcount\KZ\KZ=\refno
\autoref\newcount\Kutasov\Kutasov=\refno
\autoref\newcount\Sonoda\Sonoda=\refno
\autoref\newcount\Zwiebach\Zwiebach=\refno
\autoref\newcount\Denjoe\Denjoe=\refno.
In this approach to relativistic quantum field theory or stastical mechanics
the couplings which parameterise a theory, e.g. masses, gauge
couplings, Yukawa couplings etc., are viewed as parameters on some space $\cal
G$ and it is the geometry of this space which is studied. It is in principle
an infinite dimensional space as there are an infinite number of operators
that one can introduce into any given theory, but there are circumstances
where one might hope that studying a finite dimensional subspace may prove to
be sufficient. For example in a renormalisable field theory there are
generically only a finite number of operators that can be included in the bare
action and the properties of all other operators should be determined by these
\lq\lq basic" operators alone (for an interacting field theory these must
include composite operators). In such circumstances $\cal G$ is finite
dimensional and can be parameterised by the couplings $g^a$ associated with
each basic operator.

One is then tempted to think of $\cal G$
as being a differentiable manifold, in which case $g^a$ would be thought of as
co-ordinates with $a=1,\ldots ,n$, where $n$ is the
dimension of $\cal G$ - for example it
has been proposed that the space of couplings
for the dynamics responsible for the quantum Hall effect can
usefully  be identified with the Lobachevski
plane,\autoref\newcount\Ross\Ross=\refno.
What would be the geometrical properties of $\cal G$ in this
approach in general? For example
one might seek a consistent definition of a metric on
$\cal G$, this would give a notion of the physical
\lq\lq distance" between two theories.
A reasonable criterion for a metric is that it should be
related to the two point functions of the
theory, [\the\Denjoe] \autoref\newcount\ZamA\ZamA=\refno.
A connection  on $\cal G$ would also be important to give a rule for
transporting tensors around.
We shall see that physical amplitudes of the theory
can be thought of as tensors on
$\cal G$. Co-variant differentiation therefore would give
a rule for comparing physical amplitudes for different theories
(by this is meant theories with the same field content but different
values of the couplings). Unfortunately there is as yet no
clear physical definition of a connection, though
some suggestions have been made,
[\the\KZ] [\the\Kutasov] [\the\Sonoda] [\the\Zwiebach].
The connection is related to the operator expansion co-efficients,
and this relationship will be investigated in detail in section five.
A precise determination of the connection will not be attempted here,
however, rather the existence of
a connection will simply be be assumed and some
inferences will be drawn.

A more primitive form of differentiation also exists
in differential geometry apart from co-variant
differentiation, that of Lie differentiation, but the definition
of a Lie derivative requires choosing a vector field - it is not
intrinsic to the basic geometry of the underlying manifold.
It is more primitive in the sense that it does not require either
a metric or a connection for its definition and so does not
rely on the geometry to the same extent as the co-variant
derivative. It has been shown,
\autoref\newcount\Lassig\Lassig=\refno
\autoref\newcount\Dolan\Dolan=\refno,
that the renormalisation group equation for $N$-point amplitudes
can be thought of as an equation for Lie differentiation of
the amplitudes along the vector field defined by the $\beta$-functions
of the theory.
{}From this perspective the anomalous dimensions are seen as arising
from Lie differentiation of the basis vectors for the tangent space.
{}From a geometrical point of view the vanishing of the anomalous
dimension of the free energy is due to the fact that it is a scalar function
on $\cal G$, being the logarithm of the partition function, and is not
a tensor.

However this interpretation of the renormalisation
group equation as a Lie derivative is only valid when the points in
Euclidean space at which the $N$-point function is evaluated,
$x_1,\ldots ,x_N$, are well separated with respect to the renormalisation
length $\kappa^{-1}$.\footnote*{For simplicity we shall take
the underlying physical space in which the theory is formulated
to be $D$-dimensional Euclidean space, ${\bf R}^D$.}
It will be shown in this paper that there
are corrections to this interpretation when some of these points
get close together and that these corrections
can be expressed by the introduction
of a connection.
One of the main results presented here is a co-variant generalisation of some
formulae appearing
in reference\autoref\newcount\HughIan\HughIan=\refno
for the way in which
regularised $N$-point Green functions mix with lower
$N$-point functions under changes in the renormalisation scale $\kappa$.

We shall first state the results.
Denote the basic operators, which may be composite,
by $[\Phi_a(p)]$ (these will be defined more precisely later)
and the regularised Green functions in momentum space by
$$\eqalign{G^R(p,q)_{ab}&=<[\Phi_a(p)\Phi_b(q)]>\cr
G^R(p,q,r)_{abc}&=<[\Phi_a(p)\Phi_b(q)\Phi_c(r)]>\cr
G^R(p,q,r,s)_{abcd}&=<[\Phi_a(p)\Phi_b(q)\Phi_c(r)\Phi_d(s)]>\qquad
\hbox{etc.,}\cr}\autoeq$$
(square brackets around an operator denote that it is regularised).
Then, assuming \hbox{$<[\Phi_a(p)]>=0$} for simplicity, it will be shown that
the renormalisation group equations for three and four point functions are
$$\eqalign{
   \left[
          \Bigl(\kappa{\partial\over\kappa}\Vline\sub{g} + {\cal L}_\beta\Bigr)
          G^R(p,q,r)
   \right]_{abc}
      & = {\tau_{ab}}^d
               G^R_{dc}(p+q,r) + {\tau_{bc}}^d
               G^R_{da}(q+r,p) + {\tau_{ca}}^d
               G^R_{db}(r+p,q)
      \cr
      & \qquad\qquad\qquad\qquad\qquad\qquad+\cdots
      \cr
  \left[
           \Bigl(\kappa{\partial\over\kappa}\Vline\sub{g} + {\cal
L}_\beta\Bigr)
           G^R(p,q,r,s)
         \right]_{abcd}
       & =\left[ {\tau_{ab}}^f
           G^R_{fcd}(p+q,r,s) + \; 5\;
           \hbox{terms}
          \right]
      \cr
       &  \qquad
         - \left[ \bigl(\Del_a{\tau_{bc}}^f\bigr)
              G^R_{fd}(p+q+r,s)+ \; 3\; \hbox{terms}
          \right]  + \cdots
       \cr}
\autoeq$$\newcount\inrge\inrge=\count90
The notation ${\cal L}_\beta$ denotes the Lie derivative e.g.,
$${\cal L}_\beta{G^R_{ab}}^c=\beta^d\partial_d{G^R_{ab}}^c
+\bigl(\partial_a\beta^d\bigr){G^R_{db}}^c
+\bigl(\partial_b\beta^d\bigr){G^R_{ad}}^c
-\bigl(\partial_d\beta^a\bigr){G^R_{ab}}^d.\autoeq$$
\newcount\liedef\liedef=\count90
The matrix $\partial_a\beta^b$ is the matrix of anomalous
dimensions which mixes the operators under renormalisation.
In equation (\the\inrge) dots
denote terms which are monomials of the momenta to the
power $D$, such terms are only significant when all $N$ points of
the $N$-point function are so close to one another that
they are unresolvable on the length scale $\kappa^{-1}$.

The tensor
${\tau_{ab}}^c$ appearing in the RG equations above
is symmetric in $a$ and $b$ and
involves the second co-variant derivative of the
$\beta$-functions,
$${\tau^a}_{bc}=\Del_b\Del_c\beta^a - {R^a}_{cbd}\beta^d,\autoeq$$
\newcount\taudef\taudef=\count90
with ${R^a}_{cbd}$ the curvature associated with the connection.
This tensor is central to the treatment presented here
as it governs the way in
which $N$-point functions (which are rank $N$ co-variant tensors
in the co-tangent space $T^*({\cal G})$ at the point $g$)
mix with tensors of lower rank, it is a co-variant generalisation
of $\partial_a\partial_b\beta^c$ which appears in [\the\HughIan].

The connection can be related to the operator product
expansion co-efficients. The idea that a connection should be
related to the OPE co-efficients has been expressed before,
[\the\Kutasov] [\the\Sonoda]. In particular the analysis of the
OPE presented in section five
is similar in spirit to that of Sonoda [\the\Sonoda],
but it is implemented in a different way in that reference.
The OPE co-efficients have also been related to the second derivatives
of the $\beta$-functions, at least near a critical point, by
Zamolodchikov,
\autoref\newcount\ZamB\ZamB=\refno.
He defines quantities ${{\tilde C}^a}_{bc}$ which are related
to the OPE co-efficients ${C^a}_{bc}(x)$ and shows that
${{\tilde C}^a}_{bc}=\partial_b\partial_c\beta^a$.

The second main result of this paper is that certain OPE co-efficients satisfy
the following renormalisation group equation, in momentum space,
$$\bigl(\kappa{\partial\over\partial\kappa}\Vline\sub g +{\cal L}_\beta\bigr)
{C^R_{ab}}^c(p)={\tau_{ab}}^c + \cdots,\autoeq$$
\newcount\operg\operg=\count90
where ${C^R_{ab}}^c(p)$ are regularised OPE co-efficients,
in the sense that their integral over all $p$ (or equivalently
over all space) is finite.
This equation is a co-variant generalisation of a result presented
in\autoref\newcount\NLSR\NLSR=\refno. It relates the OPE co-efficients
to the connection via the tensor ${\tau_{ab}}^c$.

The layout of the paper is as follows. In section two the renormalisation
group equation, including the possibility of composite operators,
will be discussed from a geometrical point of view.
It will be argued that, at least when all of the points
are well separated, the equation reduces to nothing more than
the definition of the Lie derivative of tensors on $\cal G$
with respect to the vector field given by the $\beta$-functions of the theory.
In section three the necessary changes required to regulate Green functions
when two of the points
get close to one another are discussed and the technique
of using position dependent couplings is outlined. The
resulting expressions are not co-variant under general co-ordinate
transformations on $\cal G$.
Section four is devoted to the development of co-variant expressions
and a co-variant renormalisation group equation. It is shown
how the renormalisation group flow mixes up tensors of different rank.
In section five the operator product expansion co-efficients are discussed
and a co-variant renormalisation group equation for them is derived.
It is argued that the connection is related to the OPE co-efficients.
In section six the results are summarised and some comments are
made on possible future directions of development.
A derivation of the non-covariant expression for regularised $N$-point
functions
with arbitrary $N$ is given in an appendix, for massless theories
in four dimensions. A second appendix gives the co-variant renormalisation
group equation for four point functions with arbitrary momenta.
%\vskip .5cm
\vfill\eject
{\bf \S 2 The Renormalisation Group Equation}
\vskip .5cm
The regularisation of quantum amplitudes involving
composite operators is more subtle than for those involving just
elementary fields, at short distances
new divergences appear over and above those
present in the standard treatments. In reference [\the\HughIan] a technique
was described for dimensional regularisation of amplitudes
involving only those composite operators associated with
couplings appearing in the original
Lagrangian of the theory and this was extended in [\the\NLSR] to include
more general composite operators. For example if we consider the
renormalised composite operator $[\phi^4]$ in $\lambda\phi^4$ theory
in four dimensions
then the operator $[\phi^4(x)][\phi^4(y)]$ is singular as $x\rightarrow y$
and requires a new subtraction (here and subsequently square brackets
around an operator or product of operators means that it is regularised, thus
$[\phi^4(x)][\phi^4(y)]\ne[\phi^4(x)\phi^4(y)]$).
The technique adopted in [\the\HughIan] involves defining \lq basic' composite
operators, one associated with every coupling $g^{a_0}$ appearing in the bare
Langrangian density $L_0(x)$.
These basic operators are given by $\Phi_{a_0}(x)=\partial_{a_0}L_0(x)$.
For example in $\lambda\phi^4$ in four dimensions for $g^{a_0}=\lambda_0$
one has $\Phi_{\lambda_0}={1\over 4!}\phi_0^4$.
In a renormalisable
theory there are a finite number $n$ of these operators,
where $n$ is the number of couplings $a_0=1,\ldots,n$. $\Phi_{a_0}$ are
of course bare operators. Renormalised operators can be defined by
$[\Phi_a(x)]={Z_a}^{b_0}\Phi_{b_0}(x)$ where ${Z_a}^{b_0}$ is a matrix
of renormalisation co-efficients
which mixes operators. This matrix can be interpreted as a co-ordinate
transformation matrix
$${Z_a}^{b_0}={\partial g^{b_0}\over\partial g^a}\autoeq$$
\newcount\Zdef\Zdef=\count90
in which $g^a$ are renormalised couplings and
$[\Phi_a(x)]=\partial_aL_0(x)$. Thus the space of couplings $\cal G$
is viewed as a $n$-dimensional differentiable manifold with $g^a$ and $g^{a_0}$
being different co-ordinate systems on $\cal G$. The bare couplings,
$g^{a_0}(g^a,\epsilon)$,
are analytic functions of $g^a$ and of the regularisation parameter $\epsilon$,
provided $\epsilon\ne 0$
($\epsilon=D-4$ in dimensional regularisation). The matrix ${Z_a}^{b_0}$
is a co-ordinate transformation matrix.
Viewed from this geometric perspective the quantities
$$\Phi(x)=[\Phi_a(x)]dg^a=\Phi_{a_0}(x)dg^{a_0}\autoeq$$
are operator valued one-forms on the co-tangent space $T^*({\cal G})$.
This picture has also proven useful in
conformal field theories in two dimensions where the
operators $[\Phi_a]$ are primary
fields, [\the\ZamB].

$N$-point Green functions are now rank $N$ tensors on $\cal G$.
Provided all the points $x_i$ are well separated,
$$G^{(N)}_{a_1\cdots a_N}(x_1,\ldots,x_N)=
   <[\Phi_{a_1}(x_1)]\cdots[\Phi_{a_N}(x_N)]>.\autoeq$$
Note that in general the tensor $G^{(N)}_{a_1\cdots a_N}(x_1,\ldots,x_N)$ has
no
particular symmetry properties.
When all the $x_i$ are well separated the renormalisation group equation
has a very simple geometrical
interpretation, it is simply the Lie derivative of
$G^{(N)}_{a_1\cdots a_N}$ with respect to the vector field on $T({\cal G})$
given by the $\beta$-functions of the theory
$\vec{\beta}=\beta^a{\partial\over\partial g^a}$, [\the\Lassig] [\the\Dolan].
To see this we simply write the $N$-point functions with the
basis $dg^a$ for real valued one-forms included
$$\eqalign{
G^{(N)}&=
 <[\Phi_{a_1}(x_1)]\cdots [\Phi_{a_N}(x_N)]>dg^{a_1}\cdots dg^{a_N}\cr
 &=<\Phi_{a_{01}}(x_1)\cdots\Phi_{a_{0N}}(x_N)>dg^{a_{01}}\cdots dg^{a_{0N}}.
\cr}
\autoeq$$
The usual renormalisation group argument is now applied
to $G^{(N)}$, it should be independent of the renormalisation point $\kappa$.
Thus
$$\kappa{d\over d\kappa}G^{(N)}=
\bigl(\kappa{\partial\over\partial\kappa}\Vline\sub{g} +{\cal L}_\beta\bigr)
G^{(N)}=0.\autoeq$$
This immediately leads to
$$\eqalign{
\kappa{\partial\over\partial\kappa}\Vline\sub{g}
   & G^{(N)}_{a_1\cdots a_N}(x_1,\ldots,x_N)=
   \cr
   &-\beta^b\partial_bG^{(N)}_{a_1\cdots a_N}(x_1,\ldots,x_N)
    -\sum_{i=1}^N\bigl(\partial_{a_i}\beta^b\bigr)
    G^{(N)}_{a_1\cdots a_{i-1}ba_{i+1}\cdots a_N}(x_1,\ldots,x_N),
\cr}\autoeq$$
\newcount\LieRG\LieRG=\count90
where we have used
$$\kappa{d\over d\kappa}dg^a=d\Bigl(\kappa{dg^a\over d\kappa}\Bigr)
=d\beta^a=\partial_b\beta^adg^b.\autoeq$$
The matrix of anomalous dimensions,
$\partial_b\beta^a$, is thus seen to come from
Lie dragging of the basis one-forms $dg^a$.
Note that equation (\the\LieRG) is co-variant under
general co-ordinate transformations, even though the derivatives
on the right hand side are not co-variant, because the Lie
derivative is co-variant by
construction, \autoref\newcount\Kobayashi\Kobayashi=\refno.
There is no need to introduce a connection to define Lie derivatives.
However the interpretation of the matrix $\partial_a\beta^b$ as
having physical significance is tied in to a special choice of
co-ordinates. More generally one would expect a co-variant
generalisation of this matrix, $\Del_a\beta^b$,
to have the physical interpretation
of a matrix of anomalous dimensions, [\the\Lassig].

This treatment of the RG equation, though conceptually
simple, is not the whole story. We must be careful to
regularise $[\Phi_{a_i}(x_i)][\Phi_{a_j}(x_j)]$ whenever any two
of the points $x_i$ and $x_j$ start getting close to one another.
The operator product expansion co-efficients clearly play an
important role here and this combination becomes a single
renormalised composite operator as $x_i\rightarrow x_j$.
Thus the regularised Green functions,
$$G^{R(N)}_{a_1\cdots a_N}(x_1,\cdots x_N)=
<[\Phi_{a_1}(x_1)\cdots\Phi_{a_N}(x_N)]>,\autoeq$$
are linear combinations of all the lower, unregularised ones,
$G^{(M)}$ for $M\le N$ including $M=0$. This phenomenon manifests
itself at the level of the renormalisaion group by the fact
that $G^{R(N)}$ gets mixed up with tensors of lower rank under RG flow.
This mixing was exhibited in [\the\HughIan], but the tensor expressions
in that reference were not co-variant. For example the mixing
co-efficients involved the second derivative of the $\beta$-functions,
$\partial_a\partial_b\beta^c$ which is clearly not a tensor and this can
only be consistent if it is legitimate to put a flat connection
on $T({\cal G)}$ and a co-ordinate system can be found
in which the connection co-efficients vanish (e.g. if $T({\cal G})$
admits a flat metric, with $g^a$ Cartesian co-ordinates, and the
connection is the Levi-Civita connection).
The remedy for this problem is pointed out in [\the\HughIan], a connection
on $T({\cal G})$ must be introduced. It is even indicated how this
should be done, but the authors do not do it because they do not
know how to calculate the connection, in general.
However a connection must be introduced, whether it be flat or not,
in order to make the renormalisation group equation co-variant
and the approach adopted here will be to introduce one,
without any prescription as to how it might be calculated,
and co-variant expressions will be derived.

To appreciate the necessity of the aforementioned regularisation
consider the two point functions, $G_{ab}(x_1,x_2)$,
for a theory in flat
Euclidean space, ${\bf R}^D$, where $D=4-\epsilon$ and
$a$ and $b$ are indices associated with dimensionless couplings.
In dimensional regularisation $[\Phi_a]$ are dimension
$4-\epsilon$ operators. The renormalised two point Green
functions are of the
form,\footnote*{Henceforth the superscript $(N)$ on $N$-point functions will
be omitted since it is clear from the index structure on $G$
which value of $N$ is under consideration.}
\autoref\newcount\Brown\Brown=\refno
$$G^R_{ab}(x,y)=G_{ab}(x,y)+\kappa^{-\epsilon}A_{ab}\Box\Box
\delta(x-y),
\autoeq$$\newcount\GRtwo\GRtwo=\count90
\newcount\twopoint\twopoint=\count90
where $A_{ab}(g,\epsilon)$ is a tensor on $\cal G$, independent of $x$ and $y$,
but depending on $g^a$ and
containing poles in $\epsilon$ in general. $\Box$ is the four
dimensional Laplacian in Euclidean space,
$\Box=\partial^\mu\partial_\mu$.
The tensor $A_{ab}(g,\epsilon)$ is chosen to cancel the singularities
at $x\approx y$ in $G_{ab}(x,y)$, so as to render $\int d^DxG^R_{ab}(x,y)$
finite. These counterterms introduce corrections into the RG equation
which will be developed in section four. First we develope a technique
for determining regularised $N$-point functions for general $N$.
\vskip .5cm
{\bf \S 3 Regularised $N$-point Functions}
\vskip .5cm
The technique developed in [\the\HughIan] for handling the counterterms
described in the previous section
will now be summarised,
leaving out the connection on $T^*({\cal G)}$ until the next section.
Expressions for the regularised $N$-point function,
in the absence of a connection, can be derived by induction.
The renormalised Green functions are
obtained by introducing position dependent renormalised couplings,
$g^a(x)$ so that
$[\Phi_a(x)]={\delta S_0\over\delta g_a(x)}$,
where $S_0$ is the action $S_0=\int d^DxL_0(x)$ with $x^\mu$ Cartesian
co-ordinates on ${\bf R}^D$.
Next a counterterm proportional to the identity, which
involves derivatives of $g^a(x)$,
is subtracted from the bare action,
$$\tilde S_0(g,\epsilon)=S_0(g,\epsilon)
  -{\kappa^{-\epsilon}\over 2}
\int_{{\bf R}^D}d^ DxA_{ab}\Box g^a\Box g^b.\autoeq$$
\newcount\cterm\cterm=\count90

Defining the generating functional
in the usual way,
$$W=-\ln\Z\qquad\hbox{where}\qquad\Z=\int{\cal D}\varphi e^{-\tilde
S_0},\autoeq$$
\newcount\Zdef\Zdef=\count90
allows the regularised $N$-point functions to obtained
by functional
differentiation.
Thus
$$G^R_{a_1\cdots a_N}(x_1,\ldots,x_N)=(-1)^{N+1}{\delta^N W\over
\delta g^{a_1}(x_1)\cdots\delta g^{a_N}(x_N)}.\autoeq$$
\newcount\funcdif\funcdif=\count90
Now the two point functions (\the\twopoint) can be obtained from
$$G^R_{ab}(x,y)={\delta^2\ln\Z\over\delta g^a(x)\delta g^b(y)}
\Vline\sub{\partial_\mu g=0}
=\Bigl<{\delta\tilde S_0\over\delta g^a(x)}
{\delta\tilde S_0\over\delta g^b(y)}\Bigr>\Vline\sub{\partial_\mu g=0}
-\Bigl<{\delta^2\tilde S_0\over\delta g^a(x)\delta g^b(y)}\Bigr>
\Vline\sub{\partial_\mu g=0},\autoeq$$\newcount\regtwo\regtwo=\count90
where it is assumed for simplicity that
$$<[\phi_a(x)]>\vert_{\partial_\mu g^a
=0}=0.\autoeq$$\newcount\vev\vev=\count90
This assumption simplifies some of the
equations. Non-zero expectation values are easily accounted for
by using (\the\LieRG) with $N=1$ giving
$$\kappa{\partial\over\partial\kappa}\Vline\sub{g}G_a=
-\beta^b\partial_bG_a-\partial_a\beta^bG_b\autoeq$$
(one point Green functions, $G_a$, are of course independent
of position because of translational invariance).

The new counter term in (\the\cterm) gives rise to
$$\eqalign{
{\delta^2\tilde S_0\over\delta g^a(x)\delta g^b(y)}
\Vline\sub{\partial_\mu g=0}
&=\left({\delta\over\delta g^a(x)}[\phi_b(y)]\right)
\Vline\sub{\partial_\mu g=0}\cr
&=\delta(x-y){K^c}_{ab}[\phi_c(x)]
\vline\sub{\partial_\mu g=0}-
\kappa^{-\epsilon}A_{ab}\Box\Box\delta(x-y),\cr}
\autoeq$$\newcount\doeq\doeq=\count90
where it is assumed that the bare basic operators are independent
of the couplings, \break
${\delta\phi_{b_0}(y)\over\delta g_{a_0}(x)}=0$.
The quantities ${K^c}_{ab}$ here are defined by
$${K^c}_{ab}(g,\epsilon)=\partial_a\partial_b g^{d_0}
\left({\partial g^c\over\partial g^{d_0}}\right)=
\partial_a{Z_b}^{d_0}{(Z^{-1})_{d_0}}^c,\autoeq$$
\newcount\Kdef\Kdef=\count90
and contain poles in $\epsilon$.
Combining equations (\the\regtwo) and (\the\doeq) now gives the
regularised two point functions (\the\GRtwo).

In four dimensions there are other counterterms proportional
to the identity that can be added to $S_0$ which are necessary
for regularisation when $N>2$. For simplicity we shall
assume that only couplings that are dimensionless in four dimensions
appear in the Lagrangian (no masses). Including masses introduces
more terms but is straightforward in principle.

By simple dimensional
analysis only terms involving the appropriate number of derivatives
of the dimensionless
couplings can contribute. The most general counterterm,
invariant under parity transformations, consists
of the following combination (modulo integration by parts),
$$\tilde S_0(g,\epsilon)=S_0(g,\epsilon)-I_0(g,\epsilon),$$
where
$$\eqalign{
I_0(g,\epsilon)&=\int_{{\bf R}^D}d^Dx{\cal I}_0\cr
&={1\over 2}\int_{{\bf R}^D}d^DxA_{ab}\Box g^a\Box g^b
+{1\over 2}\int_{{\bf R}^D}d^DxB_{abc}\partial_\mu g^a \partial^\mu g^b
\Box g^c\cr
&\qquad\qquad\qquad
+{1\over 4}\int_{{\bf R}^D}d^DxC_{abcd}\partial_\mu g^a
\partial^\mu g^b\partial_\nu
g^c\partial^\nu g^d.\cr}\autoeq$$\newcount\ict\ict=\count90
$B_{abc}(g,\epsilon)$ and $C_{abcd}(g,\epsilon)$ are new quantities
with no explicit $x$ and $y$ dependence,
but depending on $g^a(x)$ and
containing poles in $\epsilon$. As emphasised in [\the\HughIan]
they are not tensors because they do not transform co-variantly
but this will be remedied later when a connection on
$T^*({\cal G)}$ is included.
Note the symmetries $B_{abc}=B_{bac}$ and
$C_{abcd}=C_{cdab}=C_{bacd}=C_{abdc}$.

The structure of this counterterm would be more complicated
if there were masses around, but these can be treated by similar
techniques and (\the\ict) will be sufficient for the purposes of
illustration. $I_0$ also depends on the number of
dimensions $D$, for example in two dimensions $B_{abc}$ and $C_{abcd}$
do not appear and
$$
{\cal I}_0={1\over 2}A_{ab}\partial_\mu g^a\partial^\mu g^b.\autoeq
$$
The arguments here will be illustrated using (\the\ict).
Thus, for example, the operator which gives a finite $3$-point
function is,
$$\eqalign{
{\delta\tilde S_0\over\delta g^a(x)}
{\delta\tilde S_0\over\delta g^b(y)}
{\delta\tilde S_0\over\delta g^c(z)}
-{\delta^2\tilde S_0\over\delta g^a(x)\delta g^b(y)}
&{\delta\tilde S_0\over\delta g^c(z)}
- {\delta^2\tilde S_0\over\delta g^a(x)\delta g^c(z)}
{\delta\tilde S_0\over\delta g^b(y)}\cr
&-{\delta^2\tilde S_0\over\delta g^b(y)\delta g^c(z)}
{\delta\tilde S_0\over\delta g^a(x)} +
{\delta^3\tilde S_0\over\delta g^a(x)\delta g^b(y)\delta g^c(z)},\cr}$$
and it involves $B_{abc}$ as well as derivatives of $A_{ab}$.

In momentum space the regularised two, three and four point functions
can thus be determined in terms of their unregularised counterparts
and $B_{abc}$ and $C_{abcd}$
by setting $\partial_\mu g^a=0$ in the
appropriate finite operators.
The two and three point functions are derived in reference [\the\HughIan]
and are, after Fourier transforming to momentum
space,
$$G^R_{ab}(p,q)=G_{ab}(p,q)+\kappa^{-\epsilon}A_{ab}p^2q^2,
\qquad \hbox{with}\quad
p+q=0\autoeq$$\newcount\threepoint\threepoint=\count90
$$\eqalign{
G^R_{abc}(p,q,&r)=G_{abc}(p,q,r)-{K^d}_{ab}G_{dc}(p+q,r)
-{K^d}_{bc}G_{da}(q+r,p)-{K^d}_{ca}G_{db}(r+p,q)\cr
&-\kappa^{-\epsilon}\bigl(p^2q^2A_{ab,c}+q^2r^2A_{bc,a}+
r^2p^2A_{ca,b}+r^2p.qB_{abc}+p^2q.rB_{bca}
+q^2r.pB_{cab}\bigr)\cr}$$
with $p+q+r=0$. Using the same techniques the four point function can
also be determined to be
$$\eqalign{
G^R_{abcd}(p,q,r,s)=&G_{abcd}(p,q,r,s)-\bigl({K^e}_{ab}G_{ecd}(p+q,r,s)+5
\;\hbox{permutations}\bigr)\cr
&\qquad+\bigl({K^e}_{ab}{K^f}_{cd}G_{ef}(p+q,r+s)+2
\;\hbox{permutations}\bigr)\cr
&\qquad+\bigl({K^e}_{abc}G_{ed}(-s,s)+3
\;\hbox{permutations}\bigr)\cr
&\qquad+\kappa^{-\epsilon}\bigl(r^2s^2A_{cd,ab}+5
\;\hbox{permutations}\bigr)\cr
&\qquad+\kappa^{-\epsilon}\bigl(p^2(r.s)B_{cda,b}+11
\;\hbox{permutations}\bigr)\cr
&\qquad+2\kappa^{-\epsilon}
\bigl((p.q)(r.s)C_{abcd}+2
\;\hbox{permutations}\bigr),
\cr}\autoeq$$\newcount\fourpoint\fourpoint=\count90
with $p+q+r+s=0$.
In these expressions a comma denotes
partial differentiation with respect to the couplings $g^a$
and ${K^e}_{abc}$ is defined as
$${K^a}_{bcd}=
\bigl({\partial^3 g^{f_0}\over\partial g^b\partial g^c\partial g^d}\bigr)
{\partial g^a\over\partial g^{f_0}}
=\partial_d{K^a}_{bc}+{K^e}_{bc}{K^a}_{ed}.\autoeq$$
\newcount\kthree\kthree=\count90
In equation (\the\fourpoint)
some terms involving exceptional momenta (e.g. $p+q=r+s=0$) have been
omitted for clarity.
The reason such momenta cause extra terms to arise
is that $N$-point amplitudes (\the\funcdif)
are not simple expectation values of the basic operators, for
$N\ge 4$. For example the four point function in position space,
$G_{abcd}(x,y,z,t)$,
involves products of two
point functions $G_{ab}(x,y)G_{cd}(z,t)$. However such terms are discarded
when momenta are non-exceptional because both the two point fucntions
are separately translationally invariant. Thus in momentumm space
this reads $G_{ab}(p,q)G_{cd}(r,s)\delta(p+q)\delta(r+s)$
and excluding exceptional momenta excludes such terms. This is
also true of the expectation values themselves, $<[\Phi_a(p)]>$.
Translational invariance demands that these vanish except at
$p=0$, thus if they were to appear as factors
in a $N$-point amplitude momentum conservation would require
exceptional momenta among the other $N-1$ momenta. Hence excluding
exceptional momenta automatically excludes expectation values
for the basic operators. This simplifies some of the following formulae.
In all of the ensuing expressions it will be
assumed that none of the momenta is exceptional.

The structure of the terms involving the unregularised Green
functions on the right hand side of equations (\the\threepoint) and
(\the\fourpoint) is independent of the dimension $D$ in which we
are working and is not affected by the introduction of masses.
In other dimensions, or in theories with masses, only the $A,B$ and $C$ terms
differ. e.g. in two dimensions only the $A$ term is there
(with $p.q$ instead of $p^2q^2$)
since there are no $B$ or $C$ terms in two dimensions. Also
extra terms independent of the $G^{(M)}$'s appear when there
are masses.

The various terms in equation (\the\fourpoint) can be given the
following physical interpretation. The integral of the
left hand side with respect to any of its arguments is finite and the terms
on the right hand side involving three point functions are necessary
in order
to cancel singularities in the integral of the unregularised $G_{abcd}$ that
occur whenever one of the three independent momenta gets large.
This happens if two of the four points get close in space so that
the unregularised Green function becomes effectively a three point function
multiplied by divergent operator product expansion co-efficients.
The divergent part is extracted as a $\delta$-function in
position space and
the three point functions appearing on the right hand side
of (\the\fourpoint)
cancel this
divergence in momentum space. These three point functions only depend on two
of the momenta because of overall momentum conservation,
e.g. $G_{ecd}(p+q,r,s)=G_{ecd}(-r-s,r,s)$
and independence of the third momentum corresponds to
$\delta$-function singularities in position space.
The ${K^a}_{bc}$ are thus related to the operator
product expansion co-efficients in a manner which will be
analysed more fully later.
Similarly the terms on the right
hand side involving two point functions cancel singularities
in double integrals of $G_{abcd}$
that arise when two of the momenta get large.
Lastly the terms involving momenta to the fourth power cancel
the singularities that occur when all the four points in position
space collapse to a single point in a triple integral - the
momentum structure of
these terms indicates that they correspond to fourth derivatives
of $\delta$-functions and are thus more singular than the other terms.

The regularised $N$-point functions can be obtained from this
technique by induction and are given in an appendix.
The expressions given here are however not co-variant
under co-ordinate transformations on $\cal G$.
\vskip .5cm
{\bf \S 4 A Co-variant Renormalisation Group Equation}
\vskip .5cm
The regularised $N$-point functions in the previous section
have been derived assuming the connection
vanishes. Everything can now be made co-variant by
introducing a connection as described in [\the\HughIan].
The basic idea is the following. For position dependent
couplings the matrix ${M_\mu}^a=\partial_\mu g^a$
gives a map between $T^*({\cal G})$
and $T^*({\bf R}^D)$,
$$\matrix{
{M_\mu}^a:&T^*({\cal G})&\rightarrow &T^*({\bf R}^D)\cr
          &\omega_a     &\mapsto &\quad\omega_\mu:={M_\mu}^a\omega_a,\cr}
\autoeq$$
where $\omega_a$ is a one-form on $T^*({\cal G})$ and $\omega_\mu$
a one-form on $T^*({\bf R}^D)$. Introduce a connection
${\Gamma^a}_{bc}$
on $T({\cal G})$. Then for any vector $V^a\in T({\cal G})$
a co-variant derivative $\Del_\mu$ mapping
$T({\cal G})\rightarrow T({\cal G})\otimes T^*({\bf R}^D)$
can be defined by
$$\Del_\mu V^a:=\partial_\mu V^a + {{{\cal A}_\mu}^a}_bV^b\qquad
\hbox{where}\qquad {{{\cal A}_\mu}^a}_b={M_\mu}^c{\Gamma^a}_{cb}.\autoeq$$
In this expression $\partial_\mu V^a$ is to be interpreted as
$\partial_\mu g^b\partial_b V^a$. Thus $\Del_\mu V^a={M_\mu}^b\Del_bV^a$.
Now the Laplacian in flat Euclidean
space acting on couplings is modified to read
$$\Box g^a\rightarrow\Del^2 g^a
=\Del_\mu\partial^\mu g^a=\partial_\mu\partial^\mu g^a + {\Gamma^a}_{cb}
\partial_\mu g^b \partial^\mu g^c.\autoeq$$
The idea here is that the co-ordinates
$g^a$ are not vectors, they are $n$ functions
on $\cal G$, and $\partial^\mu g^a$ are $n$ vectors on $T({\bf R}^D)$.
The co-varaint derivative
is now used in the definition of $I_0$ in (\the\ict)
to replace $\partial_\mu g^a$. Thus (\the\ict) now reads
$$\eqalign{
{\cal I}_0
={1\over 2}A_{ab}\Del^2 g^a\Del^2 g^b
+{1\over 2}B_{abc}\partial_\mu g^a \partial^\mu g^b \Del^2 g^c
+{1\over 4}C_{abcd}\partial_\mu g^a\partial^\mu g^b\partial_\nu
g^c\partial^\nu g^d\cr},\autoeq$$\newcount\cict\cict=\count90
where all three of $A_{ab}, B_{abc}$ and $C_{abcd}$ are now tensors.
The $A,B$ and $C$ in (\the\cict) are not the same
as those in (\the\ict) - they
differ by terms involving the connection. From now on these
symbols will refer exclusively to the co-variant forms in equation
(\the\cict).
Note that only the symmetric part
of the connection, ${\Gamma^a}_{cb}={\Gamma^a}_{bc}$, is relevant
and so it is sufficient for our needs to take the connection to
be symmetric.

One defines a curvature tensor in the usual way,
$${R^a}_{bcd}=\partial_{c}{\Gamma^a}_{db}-\partial_{d}{\Gamma^a}_{cb}
+{\Gamma^a}_{ec}{\Gamma^e}_{db}
-{\Gamma^a}_{ed}{\Gamma^e}_{cb}.
\autoeq$$
It would be wrong to call ${R^a}_{bcd}$ a Riemann tensor
as there is no definition of a metric on $T({\cal G})$ here and
hence no Riemannian structure, only a connection.

There is probably a natural definition of a connection for
any given theory, e.g. the Knizhnik-Zamolodchikov connection
for certain conformal theories (for which ${R^a}_{bcd}=0$ despite
global
holonomy), [\the\KZ],
but the question of a general definition will not be addressed here.
Rather it will just be assumed that one exists and no prescription
will be given for calculating it, though it will be argued
in the next section that it must be related to the
operator product expansion co-efficients. Connections
on the space of couplings in other theories are also discussed in,
[\the\KZ] [\the\Kutasov] [\the\Sonoda] [\the\Zwiebach].

The calculation of $N$-point functions proceeds in principle as before,
though the introduction of the connection causes some extra
complications. Consider the covariant analogue of (\the\doeq)
$$\eqalign{
&\left({\delta\over\delta g^a(x)}[\phi_b(y)]
-{\Gamma^c}_{ab}[\phi_c(x)]\delta(x-y)\right)
\Vline\sub{\partial_\mu g=0}\cr
&\qquad\qquad=
\delta(x-y)\bigl({K^c}_{ab}-{\Gamma^c}_{ab}\bigr)[\phi_a(x)]
\vline\sub{\partial_\mu g=0}-
\kappa^{-\epsilon}A_{ab}\Box\Box\delta(x-y).\cr}
\autoeq$$\newcount\covdoeq\covdoeq=\count90
The difference ${K^c}_{ab}-{\Gamma^c}_{ab}$ is a tensor symmetric in $a$
and $b$ which
will be denoted by ${T^c}_{ab}$. Of course ${T^c}_{ab}$ may
contain poles in $\epsilon$ since ${K^c}_{ab}$ does, although
${\Gamma^c}_{ab}$ is assumed independent of $\epsilon$
and finite.

The regularised two point functions are
the same as in equation (\the\threepoint
) but the three point function is now
$$\eqalign{
G^R_{abc}(p,q,&r)=G_{abc}(p,q,r)-{T^d}_{ab}G_{dc}(p+q,r)
-{T^d}_{bc}G_{da}(q+r,p)-{T^d}_{ca}G_{db}(r+p,q)\cr
&-\kappa^{-\epsilon}\bigl(p^2q^2A_{ab;c}+q^2r^2A_{bc;a}+
r^2p^2A_{ca;b}+r^2p.qB_{abc}+p^2q.rB_{bca}
+q^2r.pB_{cab}\bigr),\cr}\autoeq$$\newcount\covthreepoint
\covthreepoint=\count90
where a semi-colon denotes co-variant differentiation,
$A_{ca;b}=\Del_bA_{ca}$.

For the four point function (\the\fourpoint), however, the situation is much
more complicated. The order in which the second derivatives
on $A_{ab}$ is taken is important. For simplicity, we
shall restrict ourselves to the symmetric point in momentum space,
$$p_i.p_j={\mu^2\over 3}\left(4\delta_{ij}-1\right).\autoeq$$
The co-variant result is
$$\eqalign{
G^R_{abcd}(p,q,r,s)=&G_{abcd}(p,q,r,s)-\bigl[{T^e}_{ab}G_{ecd}(p+q,r,s)+5
\;\hbox{permutations}\bigr]\cr
&\qquad+\bigl[{T^e}_{ab}{T^f}_{cd}G_{ef}(p+q,r+s)+
2\; \hbox{permutations}\bigr]\cr
&\qquad+\left[
\bigl({T^e}_{bc;a}+{T^f}_{bc}{T^e}_{af}\bigr)
G_{ed}(-s,s)+\left\{\matrix{b\leftrightarrow d\cr
                            q\leftrightarrow s\cr}\right\}
            +\left\{\matrix{c\leftrightarrow d\cr
                            r\leftrightarrow s\cr}\right\}\right]\cr
&\qquad+\bigl({T^e}_{cd;b}+{T^f}_{cd}{T^e}_{bf}\bigr)
G_{ea}(-p,p)\cr
&\qquad+\kappa^{-\epsilon}\mu^4\Bigl[
{1\over 3}\bigl(({R^f}_{cba}+{R^f}_{bca})A_{fd}
+(b\leftrightarrow d)+(c\leftrightarrow d)\bigr)\cr
&\qquad+{1\over 3}({R^f}_{cdb}+{R^f}_{dcb})A_{fa}
+{1\over 2}\bigl(A_{cd;a;b}+\;11 \;\hbox{permutations}\bigr)\cr
&\qquad-{1\over 3}\bigl(B_{cdb;a}+11\; \hbox{permutations}\bigr)
+{2\over 9}\bigl(C_{abcd}+2 \;\hbox{permutations}\bigr)\Bigr].\cr}\autoeq$$
\newcount\covfourpoint\covfourpoint=\count90
Note that the indices $a$ and $b$ now occur on a different footing to
$c$ and $d$. The expression is not so  symmetric as (\the\fourpoint)
because co-variant derivatives do not commute.

The renormalisation group equation for regularised $N$-point
functions will now be derived, for $N=2,3,4$.
The simple
result (\the\LieRG),
for the case when the points $x_1,\ldots ,x_N$
are well separated,
is modified by the mixing with the lower $M$-point functions
($M\le N$).

The derivation involves $\beta$-functions for the counterterms
(\the\ict),
$$\eqalign{
  \kappa{d\over d\kappa}{\cal I}_0
    &=\left(\kappa{\partial\over\partial\kappa}\Vline\sub{g}+\beta^a\partial_a
      \right){\cal I}_0\cr
    &=-{1\over 2}\chi_{ab}\Del^2 g^a\Del^2 g^b -
    {1\over 2}\chi_{abc}\partial_\mu g^a\partial^\mu g^b\Del^2 g^c
    -{1\over 4}\chi_{abcd}\partial_\mu g^a\partial^\mu g^b
    \partial_\nu g^c\partial^\nu g^d,
\cr}\autoeq$$
where $\chi_{ab},\chi_{abc}$ and $\chi_{abcd}$ are finite functions of the
renormalised couplings $g$, independent of $\epsilon$.
Using the expression (\the\cict) for ${\cal I}_0$
leads to the following equations for the $\chi$'s in terms of
$A_{ab},B_{abc}$ and $C_{abcd}$,
$$\eqalign{
  \chi_{ab}=
   &\epsilon A_{ab}-\bigl({\cal L}_\beta A\bigr)_{ab}\cr
\chi_{abc}=
   &\epsilon B_{abc}-\bigl({\cal L}_\beta B\bigr)_{abc}
     -2\bigl(\Del_a\Del_b\beta^d-{R^d}_{bae}\beta^e\bigr) A_{dc}\cr
\chi_{abcd}=
   &\epsilon D_{abcd}-\bigl({\cal L}_\beta D\bigr)_{abcd}
-B_{abe}\left(\Del_c \Del_d \beta^e
-{R^f}_{dce}\beta^e\right)
-B_{cde}\left(\Del_a \Del_b \beta^e -{R^e}_{baf}\beta^f\right).
\cr}\autoeq$$\newcount\chidef\chidef=\count90
Again the symbol ${\cal L}_\beta$
here denotes Lie differentiation with respect to the vector field
$\beta$. The combination $\Del_a \Del_b \beta^c -{R^c}_{baf}\beta^f$
will occur so frequently in the sequel that it will be convenient
to define ${\tau_{ab}}^c=\Del_a \Del_b \beta^c -{R^c}_{baf}\beta^f$.
${\tau_{ab}}^c$ is a tensor symmetric in the indices $a$ and $b$.

The RG equation for $N$-point
functions now follows by application of the operator
$$\kappa{d\over d\kappa}=\kappa{\partial\over \partial\kappa}\Vline\sub g
+ {\cal L}_\beta
\autoeq$$
to the regularised $N$-point functions.
A considerable simplification is introduced
by noting that $\kappa{d\over d\kappa}$
acting on unregularised Green
functions gives zero,
$\kappa{d\over d\kappa}G^{(N)}=0$ (equation (\the\LieRG)).
Another useful identity in the derivation is
$${\bigl({\cal L}_\beta T\bigr)^a}_{bc}=-
{\tau_{bc}}^a.
\autoeq$$\newcount\lieT\lieT=\count90
Note that this expression
is finite although ${T^a}_{bc}$ itself is not.
Equation (\the\lieT) is the co-varaint generalisation
of equation (2.22) in reference [\the\HughIan].

For two and three point functions the renormalisation group equation
is obtained by applying $\kappa{d\over d\kappa}$ to
equation (\the\covthreepoint) and using (\the\chidef). The result is
$$\eqalign{
 \left[
        \Bigl(\kappa{\partial\over\kappa}\Vline\sub{g} + {\cal L}_\beta\Bigr)
        G^R(p,q)
 \right]_{ab}
      &=-\kappa^{-\epsilon}(p.q)\chi_{ab}
 \cr
   \left[
          \Bigl(\kappa{\partial\over\kappa}\Vline\sub{g} + {\cal L}_\beta\Bigr)
          G^R(p,q,r)
   \right]_{abc}
      &=\left[ {\tau_{ab}}^d
               G^R_{dc}(p+q,r) + \; 2\; \hbox{permutations}
        \right]
 \cr
      &+\kappa^{-\epsilon}\bigl(p^2q^2\Del_c\chi_{ab} + r^2 p.q
\chi_{abc}
      + \; 2\; \hbox{permutations}\bigr).
 \cr}
\autoeq$$\newcount\twothreerg\twothreerg=\count90
The RG equation for four point
functions be obtained from (\the\covfourpoint) in a similar fashion.
The result for general momenta is however rather long
and for simplicity it is again given only at the
symmetric point,
$$\eqalign{
 &\left[
        \Bigl(\kappa{\partial\over\kappa}\Vline\sub{g} + {\cal L}_\beta\Bigr)
        G^R(p,q,r,s)
  \right]_{abcd}
 =\left[ {\tau_{ab}}^f
         G^R_{fcd}(p+q,r,s) + \; 5\;
         \hbox{terms}
  \right]
  \cr
     &\qquad-\left[\bigl(\Del_a{\tau_{bc}}^f\bigr)
                G^R_{fd}(-s,s)+
         \left\{\matrix{b\leftrightarrow d\cr
                        q\leftrightarrow s\cr}\right\} +
         \left\{\matrix{c\leftrightarrow d\cr
                        r\leftrightarrow s\cr}\right\}\right]
          -\bigl(\Del_b{\tau_{cd}}^f\bigr)G^R_{fa}(-p,p)
  \cr
     &\qquad-\kappa^{-\epsilon}\mu^4
     \Bigl[
           {1\over 2}\Bigl(\Del_a\Del_b\chi_{cd}+ \; 11\;
           \hbox{terms}\Bigr)
 \cr
    &\qquad\qquad
     +{1\over 3}\Bigl(\bigl\{\bigl({R^e}_{bca}+{R^e}_{cba}\bigr)\chi_{ed}
     + (b\leftrightarrow d)+(c\leftrightarrow d)\bigr\}
     +\bigl({R^e}_{cdb}+{R^e}_{dcb}\bigr)\chi_{ea}\Bigr)
  \cr
     &\qquad\qquad-{1\over 3}
        \bigr(\Del_a{\chi_{bcd}} + 11\;\hbox{permutations}\bigr)
        +{2\over 9}\bigl(\chi_{abcd}
        + 2\;\hbox{permutations}\bigr)
     \Bigr]
  \cr
}\autoeq$$\newcount\fourrg\fourrg=\count90
Again the general form of the terms involving the $G^R$'s will be the
same for any theory, only the structure of the $\mu^4$ terms will
be different in different theories. The general expression, away
from the symmetric point is given in appendix two.

\vskip .5cm
{\bf \S 5 Operator Product Expansion Co-efficients}
\vskip .5cm

The fact that, for a given $N$, the renormalisation group flow induces mixing
with Green functions of lower order is intimately related to the
operator product expansion (OPE). The
connection between the OPE co-efficients and
${K^a}_{bc}$ was mentioned previously and this will now be
made more explicit. In particular the RG equation obeyed
by the OPE co-efficients will be shown to involve
the tensor ${\tau_{ab}}^c$.

In general the OPE involves an infinite
number of operators and the basis $[\Phi_a(x)]$ should be
extended to include higher dimension
operators,
$$[\Phi_a(x)][\Phi_b(y)]={C_{ab}}^A(x-y)\bigl[O_A\bigr({x+y\over 2}
\bigr)\bigr],\autoeq$$
\newcount\OPE\OPE=\count90
where $[O_A(x)]$ are a complete set of operators, in general
an infinite set, but certainly containing all of the
$[\Phi_a(x)]$ as a subset. ${C_{ab}}^A(x-y)$ are the OPE expansion
co-efficients, which of course are singular as $x\rightarrow y$.
If the (mass) dimensions of the operators
$[\Phi_a],[\Phi_b]$ and $[O_A(x)]$ are $d_a,d_b$
and $d_A$ respectively (including anomalous
dimensions) then dimensional counting gives
the short distance behaviour for the OPE co-efficients as
$${C_{ab}}^A(x-y)\approx \vert x-y\vert^{d_A-d_a-d_b}.\autoeq$$
Thus the most singular behaviour, for given $a$ and $b$,
is for operators
on the right hand side with the smallest values
of $d_A$. Using naive dimensions the operators of lowest dimension are
precisely
those that appear in the original bare Lagrangian,
and the same conclusion will hold for the
full dimensions provided none of the anomalous dimensions is too
large. A large anomalous dimension would probably be indicative of having
chosen unphysical degrees of freedom in the original Lagrangian.
Thus, for example, in scalar $\lambda\varphi^4$
theory in four dimensions the operators $\varphi^p_0$ for $p=1,2,3,4$
are allowed to appear in the Lagrangian but higher powers of $p$
would give a non-renormalisable theory and are excluded.
The most singular terms in the OPE expansion are therefore
given by
$$[\Phi_a(x)][\Phi_b(y)]={C_{ab}}^c(x-y)\bigl[\Phi_c\bigl({x+y\over 2}\bigr)
\bigr]
\quad+\quad\hbox{less singular terms}.\autoeq$$\newcount\lst\lst=\count90
The less singular terms can be investigated by using
the non-linear source source renormalisation techniques of
reference [\the\NLSR].
It is clear that ${C_{ab}}^c(x-y)$ are tensors on $\cal G$.

Consider the unregularised $N$-point functions in position space
$G_{a_1\cdots a_N}(x_1,\ldots ,x_N)$ when only two of the points
get close but all of the others remain well separated, e.g.
$x_1\approx x_2$. Close here means that
$\vert x_1-x_2\vert\approx\kappa^{-1}$ where $\kappa$ is the
renormalisation point. From the above discussion
we have
$$G_{a_1\cdots a_N}(x_1,\ldots ,x_N)=
{C_{a_1a_2}}^d(x_1-x_2)G_{da_3\cdots a_N}
\bigl({x_1 +x_2\over 2},x_3,\ldots ,x_N\bigr)\;
+\;\vtop{\hbox{less singular}\hbox{terms}}.
\autoeq$$\newcount\ope\ope=\count90
The purpose of the counterterms is to tame the singularity as
$x_1\approx x_2$. Referring to the regularised three point
functions, (\the\covthreepoint),
in position space,
$$\eqalign{
&G^R_{abc}(x,y,z)=\cr
&G_{abc}(x,y,z)-\delta(x-y){T^d}_{ab}G_{dc}(x,z)
-\delta(y-z){T^d}_{bc}G_{da}(y,x)-\delta(z-x){T^d}_{ca}G_{db}(z,y)\cr
&\hskip 10cm +\cdots\cr}\autoeq$$\newcount\threex\threex=\count90
it is clear that we want the combination
$\int d^Dx_1{C_{a_1a_2}}^d(x_1-x_2)-{T^d}_{a_1a_2}$
to be finite.
To this end we shall define a new tensor
$${C^R_{a_1a_2}}^c(x)=
{C_{a_1a_2}}^d(x)-
\delta(x){T^d}_{a_1a_2},\autoeq$$\newcount\regc\regc=\count90
whose integral over all space is finite. Thus
$\int d^Dx {C^R_{a_1a_2}}^d(x)$ is finite, whereas
$\int d^Dx {C_{a_1a_2}}^d(x)$ is not.

Now recall the definition of the tensors ${T^a}_{bc}$,
$${T^a}_{bc}={K^a}_{bc}-{\Gamma^a}_{bc},\autoeq$$
\newcount\Tdef\Tdef=\count90
where ${K^a}_{bc}$ is given in terms of the renormalisation
matrix ${Z^{a_0}}_b$ in equation (\the\Kdef).
Equation (\the\Tdef) can be
inverted to give an expression for ${\Gamma^a}_{bc}$ in terms of
computable quantities and ${C^R_{bc}}^a$,
$${\Gamma^a}_{bc}={K^a}_{bc}
+\int d^Dx\bigl({C^R_{bc}}^a(x)-{C_{bc}}^a(x)\bigr).
\autoeq$$\newcount\connect\connect=\count90
This equation is similar to the definition of a connection used
by Sonoda, [\the\Sonoda], except that the $K$-terms are not present
in that work since it assumed there that the basic operators
are independent of the couplings. This has the consequence
that the regularised OPE co-efficients defined in [\the\Sonoda]
are not tensors, instead they transform inhomogeneously
under general co-ordinate transformations. The $K$-terms in (\the\connect)
are present because the renormalised basis operators do depend
on the couplings, in general.
Note that equation (\the\connect) does not determine the connection
but merely expresses it in terms of the undetermined finite tensor
$\int d^Dx{C^R_{bc}}^a(x)$.

More generally one could define a position
dependent connection by smearing out the $\delta$-functions
$\delta(x){K^a}_{bc}\rightarrow {K^a}_{bc}(x)$ and defining
$${\Gamma^a}_{bc}(x)={K^a}_{bc}(x)
+{C^R_{bc}}^a(x)-{C_{bc}}^a(x).
\autoeq$$\newcount\connect\connect=\count90
Such a position dependent connection appears in the version
of the RG equation presented in [\the\KZ].

Returning to equation (\the\connect), combining
(\the\ope), (\the\threex) and (\the\regc)
now leads to a regularised version of (\the\ope),
$$G^R_{a_1\cdots a_N}(x_1,\ldots ,x_N)=
{C^R_{a_1a_2}}^d(x_1-x_2)G_{da_3\cdots a_N}
\bigl({x_1+x_2\over 2},x_3,\ldots ,x_N\bigr)\quad
+\cdots\autoeq$$\newcount\rope\rope=\count90
where the dots denote terms that are negligible provided that
none of the $x_i$ is close to $x_1$ or $x_2$ for $i\ge 3$.
We now follow the standard argument that the OPE co-efficients
also satisfy a RG equation. To this end consider the
action of
$\kappa{d\over d\kappa}
  =\kappa{\partial\over\partial\kappa}+{\cal L}_\beta$
on (\the\rope) when $N=4$,
$$
\bigl(\kappa{\partial\over\partial\kappa}\Vline\sub{g}+{\cal L}_\beta\bigr)
G^R_{abcd}(x,y,z,t)=
  \Bigl[
     \bigl(\kappa{\partial\over\partial\kappa}\Vline\sub{g}+{\cal
L}_\beta\bigr)
     {C^R_{ab}}^e(x-y)
  \Bigr]
G_{ecd}\Bigl({x+y\over 2},z,t\Bigr) + \cdots\autoeq$$
where equation (\the\LieRG) with $N=3$ has been used.
In momentum space this reads
$$
\bigl(\kappa{\partial\over\partial\kappa}\Vline\sub{g}+{\cal L}_\beta\bigr)
G^R_{abcd}(p,q,r,s)=
  \Bigl[
     \bigl(\kappa{\partial\over\partial\kappa}\Vline\sub{g}+{\cal
L}_\beta\bigr)
     {C^R_{ab}}^e\Big({p-q\over 2}\Bigr)
  \Bigr]
G^R_{ecd}(p+q,r,s)+\cdots\autoeq$$
where we have replaced the three point function on the right hand
side with its regularised counterpart - the difference only affects
the omitted terms.
Now it is clear from (\the\fourrg)
that the renormalisation group equation for $x\approx y$
and all other points well seperated
takes the form
$$\bigl(\kappa{\partial\over\partial\kappa}\Vline\sub{g}+{\cal L}_\beta\bigr)
G^R_{abcd}(p,q,r,s)=
{\tau_{ab}}^eG^R_{ecd}(p+q,r,s) +\cdots .\autoeq$$
Thus we deduce that, for large momenta,
$$\bigl(\kappa{\partial\over\partial\kappa}\Vline\sub{g}+{\cal L}_\beta\bigr)
{C^R_{ab}}^c(p)={\tau_{ab}}^c+\cdots,\autoeq$$
where the dots refer to terms that fall off with momentum,
the term exhibited on the right hand side is the most significant
at small distances.
A similar equation for the singular OPE co-efficients ${C_{ab}}^c$
is presented in reference [\the\NLSR],
but with vanishing connection so
${\tau_{ab}}^c$ reduces
to $\partial_a\partial_b\beta^c$.

The second derivative of the $\beta$ function
also appears in the treatment of the OPE by Zamolodchikov [\the\ZamB],
where a Taylor expansion of the $\beta$-functions
near a conformal theory is performed and it is shown that
the OPE co-efficients are essentially the quadratic terms
in this expansion.
For a conformal field theory Zamolodchikov shows that
$${C_{ab}}^c(x-y)={{\tilde C}_{ab}}^c{1\over\vert x-y
\vert^{d_a+d_b-d_c}}.\autoeq$$\newcount\cft\cft=\count90
$d_a,d_b$ and $d_c$ here are the dimensions (including anomalous
dimensions) of the operators concerned and ${{\tilde C}_{ab}}^c$
are independent of $\vert x-y\vert$.
Zamolodchikov argues that the basis operators can be chosen so that
$${{\tilde C}_{ab}}^c={\partial_{a}\partial_{b}\beta}^c.\autoeq$$
\newcount\cddb\cddb=\count90

His argument assumes that a metric exists and that the connection
is Levi-Civita. Riemann normal co-rdinates,
compatible with (\the\cft),
can then be chosen so that the connection vanishes
and (\the\cddb) ensues. Clearly this arguement cannot always
be applied. Even if co-ordinates can be chosen so that
the connection vanishes, it is not true that derivatives
of the connection vanish, unless the space is flat, and
these are important when more than one derivative is taken.
It would seem that the correct tensor to use is ${\tau_{ab}}^c$
rather than ${\partial_{a}\partial_{b}\beta}^c$ unless one has
reasons to believe that the curvature vanishes. It may be that
a flat connection is reasonable for fixed points (i.e. the
curvature vanishes at fixed points)
but this is not yet clear
and, even if this subsequently proves to be the case, it seems
unlikely to be true away from fixed points.

An important point of physics in the analysis presented
in this section is that the definition of
the regularised OPE co-efficients (\the\regc) requires integrating
over all of space and for {\it large} separations the less
singular terms in equation (\the\lst) may become important.
In other words the assumption that the operators $[\Phi_a]$
give the most important contributions in the general OPE (\the\OPE)
might not hold for large separations and other operators might
become significant for describing the physical degrees
of freedom of the theory at larger scales. Such a phenomenon occurs in
QCD, for example, where quarks and gluons are believed to
be the physical degrees of freedom at short distances
whereas mesons and hadrons are more appropriate for larger
scales. If one tries to integrate gluonic degrees of freedom
over all space one is hit by the infra-red problem.
In perturbation theory, at least, this would present
insurmountable problems. One must therefore include
an infra-red cut-off, for example integrating over
only a finite volume, and hope that
the volume can be made large enough that finite
volume effects are not important, but that $[\Phi_a]$
still give the most important contribution to the OPE
within the whole volunme. Such a procedure, if valid,
allows the determination of at least the short distance
behaviour of the theory using the techniques here, but
it must be borne in mind that it may not always give
sensible answers.
\vskip .5cm
{\bf \S 6 Conclusions}
\vskip .5cm
In conclusion it has been argued that $N$-point amplitudes, $G^{(N)}$,
should be thought of as tensors on the space of
couplings, $\cal G$, and the renormalisation group equation mixes up
tensors of different rank, $G^{(N)}$ being related to linear
combinations of $G^{(M)}$ with $M\le N$. The crucial quantity
that determines this mixing is a tensor given by the second co-variant
derivative of the $\beta$-functions of the theory,
$${\tau^a}_{bc}=\Del_b\Del_c\beta^a - {R^a}_{cbd}\beta^d.
\eqno(\the\taudef)$$
The RG equations for two, three and four point functions for
a massless theory in four dimensions are given in equations
(\the\twothreerg) and (\the\fourrg). When all the points
are well separated in space it reduces to the definition of a Lie
derivative with respect to the vector field given by the $\beta$-functions
of the theory, equation (\the\LieRG). However when any of the points
start to get close to one another, relative to the renormalisation
length $\kappa^{-1}$, there are extra contributions.
The form of the terms involving the tensors
$\chi$ (defined in equation (\the\chidef)) is specific to massless theories
in four dimensions and result from a subtraction which is necessary
in the circumstance when all the spatial points in the Green function
are degenerate. The other terms on the right hand side of
(\the\twothreerg) and (\the\fourrg) are present in any theory and
the $M$-point functions with $M<N$ reflect singularities that arise when
some of the points start getting close to one another.
It should be observed that the mixing between tensors of
different rank is {\it linear}. This is only true for the Green
functions of the theory (excluding exceptional momenta).
Were one to consider the composite
operator analogues of the proper vertices, $\Gamma^{(N)}$,
then the resulting mixing is {\it non-linear} even when
exceptional momenta are excluded, see reference
[\the\NLSR].

No prescription as to how the connection might be calculated
in general
has been given, it is merely assumed that one must exist,
but it has been argued that it should be related to the operator product
expansion co-efficients, through equation (\the\connect),
$${\Gamma^a}_{bc}={K^a}_{bc}+
\int d^Dx\bigl({C^R_{bc}}^a(x)-{C_{bc}}^a(x)\bigr),
\eqno(\the\connect)$$
where ${K^a}_{bc}=\partial_a\partial_b g^{d_0}
\left({\partial g^c\over\partial g^{d_0}}\right)$, equation (\the\Kdef).
In this expression ${C^R_{bc}}^a(x)$ is a regularised OPE
co-efficient whose integral over all space is finite.
If one could calculate ${\Gamma^a}_{bc}$ then one
would immediately know $\int d^Dx{C^R_{bc}}^a(x)$ and vice versa.
This will not be attempted here but is clearly an interesting
programme with much scope for development.

Further questions concerning the nature of the connection present
themselves. Would it be metric compatible, if one were to give
a physically reasonable definition of a Riemannian metric on
$\cal G$? For example the Zamolodchikov metric constructed
from the two point functions of the theory
$$g_{ab}=G_{ab}(x,y,)\vline_{\vert x-y\vert = \kappa^{-1}}
\autoeq$$
might be a candidate. It is expected to be positive definite
for unitary theories. It is not clear if the Levi-Civita
connection associated with this definition of a metric would
provide useful physical information for a theory, or perhaps
it would have to be supplemented by more structure. As
mentioned earlier, the connection is symmetric so if
it is not Levi-Civita then it cannot be metric compatible -
the extra structure is not simply a torsion tensor, it
would be given by the regularised OPE co-efficients,
$\int d^Dx{C^R_{bc}}^a(x)$.

Lastly it should be emphasised that everything that has been
presented here is in terms of the local geometry, the global
structure of $\cal G$ has not beem addressed at all, but
clearly it would be very interesting to be able to ascertain
something about it.

It is a pleasure to thank Denjoe O'Connor
for stimulating discussions
on the renormalisation group, and also Prof. N.~Dragon for his
hospitality at the Institut f\"ur Theoretische Physik, Hanover
where this investigation was begun.
\vskip .5cm
{\bf Appendix 1}
\vskip .5cm
For completeness we include the non-covariant expression for the regularised
$N$-point Green functions in terms of their unregularised
counterparts, for a massless theory in four dimensions. Unregularised
in this context does not mean bare - it is always assumed that renormalised
operators $[\Phi_a(x)]$ are used in all Green functions - rather
it means regularisation in the sense of regularisation of the
infinities that occur when two or more points get close together in
the Green function.

The regularised Green functions are obtained by functionally
differentiating the generating functional,
$$\e^{-W[g^a]} = \int{\cal D}\varphi\e^{-\tilde S_0(\varphi,g^a)}.
\eqno(\the\Zdef)
$$
Thus
$$G^R_{a_1\cdots a_N}(x_1,\ldots,x_N)=(-1)^{N+1}{\delta^N W\over
\delta g^{a_1}(x_1)\cdots\delta g^{a_N}(x_N)}.
\eqno(\the\funcdif)
$$
The regularised $N$-point function can be obtained by induction.
We first write down the formula for the $N$-point function
in momentum space. It reads
$$\eqalign{
   &G^R_{a_1\cdots a_N}(p_1,\ldots,p_N)\vert_{\partial_\mu g=0}=
   \sum_{s=0}^{[N/2]}
    \sum_{\hbox{partitions}}
    {(-1)\over s!}^{N+s+r_0}
    K^{m_1}_{\pi_1}\cdots K^{m_s}_{\pi_s}
 \cr
   &\hskip 4cm \times
   G_{\pi_0m_1\cdots m_s}\Bigl(p_{\pi_0(1)},\ldots ,p_{\pi_0(r_0)},
   \Sigma_{k=1}^{r_1}p_{\pi_1(k)},\ldots,
   \Sigma_{k=1}^{r_s}p_{\pi_s(k)}\Bigr)
 \cr
   &\qquad+(-1)^N{\kappa^{-\epsilon}\over 2(N-2)!}\sum_{\hbox{permutations}}
   \partial^{(N-2)}_{\{a_3\cdots a_N}A_{a_1a_2}p^2_{a_1}p^2_{a_2\}}
 \cr
   & \qquad+(-1)^N{\kappa^{-\epsilon}\over 2(N-3)!}\sum_{\hbox{permutations}}
   \partial^{(N-3)}_{\{a_4\cdots a_N}B_{a_1a_2a_3}
   \bigl(p_{a_1}.p_{a_2}\bigr)p^2_{a_3\}}
 \cr
   &\qquad+(-1)^N{\kappa^{-\epsilon} \over (2!)^2(N-4)!}
   \sum_{\hbox{permutations}}
   \partial^{(N-4)}_{\{a_5\cdots a_N}C_{a_1a_2a_3a_4}
   \bigl(p_{a_1}.p_{a_2}\bigr)\bigl(p_{a_3}.p_{a_4\}}\bigr),
\cr}\autoeq  $$\newcount\Npointp\Npointp=\count90
where the sum over partitions involves splitting
$a_1,\ldots,a_N$ up into $s+1$ sets, $\pi_0,\ldots,\pi_s$ each
with $r_j$ elements $a_{\pi_j(1)},\ldots,a_{\pi_j(r_j)}$
such that ${\sum}^s_{j=0}r_j=N$, $0\le r_0\le N$ and $2\le r_j\le N$
for $1\le j\le s$. Thus $\pi_j=\{a_{\pi_j(1)},\ldots,a_{\pi_j(r_j)}\}$
is some subset of $a_1,\ldots,a_N$ consisting
of $r_j$ elements. The number of sets lies between $0$ and $[N/2]$
where $[N/2]$ is the integral part of $N/2$.
In any given partition
each subset occurs only once, regardless of
the ordering of its elements. For example, for $N=4$, $s$ has three
possible values, $0,1$ or $2$, and the partitions are
$$s=0:\qquad\pi_0=\{a_1,a_2,a_3,a_4\}$$
$$s=1:\quad
\left\{\matrix{\pi_0=\{a_1,a_2\}\;\pi_1=\{a_3,a_4\}, &\qquad
         \pi_0=\{a_1,a_3\}\;  \pi_1=\{a_2,a_4\}, \cr
         \pi_0=\{a_1,a_4\}\;  \pi_1=\{a_2,a_3\}, &\qquad
         \pi_0=\{a_2,a_3\}\;  \pi_1=\{a_1,a_4\}, \cr
         \pi_0=\{a_2,a_4\}\;  \pi_1=\{a_1,a_3\}, &\qquad
         \pi_0=\{a_3,a_4\}\;  \pi_1=\{a_1,a_2\}, \cr
         \pi_0=\{a_1\}\;      \pi_1=\{a_2,a_3,a_4\}, &\qquad
         \pi_0=\{a_2\}\;      \pi_1=\{a_1,a_3,a_4\}, \cr
         \pi_0=\{a_3\}\;      \pi_1=\{a_1,a_2,a_4\}, &\qquad
         \pi_0=\{a_4\}\;      \pi_1=\{a_1,a_2,a_3\}, \cr
         \pi_0=\emptyset\;   \pi_1=\{a_1,a_2,a_3,a_4\} \cr}\right.
$$
$$s=2:\quad
\left\{\matrix{
        \pi_0=\emptyset \;\pi_1=\{a_1,a_2\} \;\pi_1=\{a_3,a_4\}, &\qquad
        \pi_0=\emptyset \;\pi_1=\{a_1,a_3\} \;\pi_2=\{a_2,a_4\}, \cr
        \pi_0=\emptyset \;\pi_1=\{a_1,a_4\} \;\pi_2=\{a_2,a_3\}, &\qquad
        \pi_0=\emptyset \;\pi_1=\{a_2,a_3\} \;\pi_2=\{a_1,a_4\}, \cr
        \pi_0=\emptyset \;\pi_1=\{a_2,a_4\} \;\pi_2=\{a_1,a_3\}, &\qquad
        \pi_0=\emptyset \;\pi_1=\{a_3,a_4\} \; \pi_2=\{a_1,a_2\}.
\cr}\right.
\autoeq$$

The co-efficients $K^{m_j}_{\pi_j}=K^{m_j}_{a_{\pi_j(1)}\cdots a_{\pi_j(r_j)}}$
in equation (\the\Npointp) are defined analogously to (\the\Kdef),
$$K^{m}_{a_{\pi_j(1)}\cdots a_{\pi_j(r_j)}}
=\bigl(\partial_{a_{\pi_j(1)}}\cdots\partial_{a_{\pi_j(r_j)}} g^{d_0}\bigr)
\left({\partial g^m\over\partial g^{d_0}}\right).\autoeq$$
\newcount\NKdef\NKdef=\count90
Terms involving exceptional momenta are omitted
from the above expression.
It is straightforward to show that
 equation (\the\Npointp) reproduces the regularised two, three and four point
functions in the text (equations (\the\threepoint) and (\the\fourpoint)),
provided $G_a(p)=<\Phi_a(p)>=0$.

It will be more useful to work in position space in order to construct
an inductive proof. Equation (\the\Npointp) translates as
$$\eqalign{
   &G^R_{a_1\cdots a_N}(x_1,\ldots, x_N)\vert_{\partial_\mu g=0}=
   \sum_{s=0}^{[N/2]}
    \sum_{\hbox{partitions}}
    {(-1)\over s!}^{N+s+r_0}
    K^{m_1}_{\pi_1}\cdots K^{m_s}_{\pi_s}\cr
   &\qquad\qquad\qquad\qquad\times G^{(r_0+s)}_{\pi_0 m_1 \cdots m_s}
   (x_{\pi_0(1)},\ldots ,x_{\pi_0(r_0)},x_{\pi_1(1)},\ldots ,x_{\pi_s(1)})
   \prod_{j=1}^s\prod_{m=2}^{r_j}\delta_{x_{\pi_j(1)},x_{\pi_j(m)} }
 \cr
   &\qquad+{(-1)^N\kappa^{-\epsilon}\over 2(N-2)!}\Bigl[
   \partial^{(N-2)}_{a_3\cdots a_N}A_{a_1a_2}
   \delta^{\prime\prime}_{x_{a_1},x_{a_N} }
   \delta^{\prime\prime}_{x_{a_2},x_{a_N} }
   \prod_{j=3}^{N-1}\delta_{x_{a_j},x_{a_N} }\;+\;\hbox{permutations}\Bigr]
 \cr
   & \qquad+{(-1)^N\kappa^{-\epsilon}\over 2(N-3)!}\Bigl[
   \partial^{(N-3)}_{a_4\cdots a_N}B_{a_1a_2a_3}
   \bigl(
   \delta^\prime_{x_{a_1},x_{a_N}}.\delta^\prime_{x_{a_2},x_{a_N}}\bigr)
   \delta^{\prime\prime}_{x_{a_3},x_{a_N}}
   \prod_{j=4}^{N-1}\delta_{x_{a_j},x_{a_N}}\,+\,\hbox{permutations}\Bigr]
 \cr
   &\qquad+{(-1)^N\kappa^{-\epsilon} \over (2!)^2(N-4)!}\Bigl[
   \partial^{(N-4)}_{a_5\cdots a_N}
   C_{a_1a_2a_3a_4}\bigl(
   \delta^\prime_{x_{a_1},x_{a_N}}.
   \delta^\prime_{x_{a_2},x_{a_N}}\bigr)
   \bigl(
   \delta^\prime_{x_{a_3},x_{a_N}}.
   \delta^\prime_{x_{a_4},x_{a_N}}\bigr)
   \prod_{j=5}^{N-1}\delta_{x_{a_j},x_{a_N}}
 \cr
   &\hskip 10cm +\;\hbox{permutations}\Bigr],
 \cr}\autoeq  $$\newcount\Npointx\Npointx=\count90
where the $\delta$-function notation used here is a shorthand for
$\delta_{x_i,x_j}=\delta(x_i-x_j)$ and a prime denotes differentiation
with respect to the first argument of the $\delta$-function. Thus
$\delta^{\prime\prime}_{x_i,x_j}=\Box_{x_i}\delta(x_i-x_j)$ and
$\delta^\prime_{x_i,x_j}.\delta^\prime_{x_i,x_k}=\partial_{x_i}^\mu\delta(x_i-
x_j)\partial_{x_i\mu}\delta(x_i-x_k)$. The number of arguments
in the unregularised Green
functions on the right hand side has been shown explicity as a superscript
in order to try to make the formulae easier to interpret - thus
$G^{(r_0+s)}$ is a $(r_0+s)$-point function.
The fact that exceptional momenta are being excluded is interpreted in
position space as meaning that terms which factorise into
products of amplitudes which are separately translationally
invariant are omitted
from (\the\Npointx).

Proceeding inductively, we relax the condition
$\partial_\mu g^a=0$ in the regularised Green functions and
functionally differentiate (\the\Npointx)
with respect to $g^{a_{N+1}}(x_{N+1})$ and then
check that the resulting expression
agrees with (\the\Npointx) with $N$ replaced by $N+1$.
The regularised $N+1$-point function is thus given by
$$-{\delta G^R_{a_1\cdots a_N}(x_1,\ldots,x_N)\over\delta g^{a_{N+1}}(x_{N+1})}
\Vline\sub{\partial_\mu g=0}.\autoeq$$\newcount\Npopoint\Npopoint=\count90
Consider, therefore, a generic term from the right hand
side of (\the\Npointx),
$$F_s[g^a(x)]:=
(-1)^{N+s+r_0}
    K^{m_1}_{\pi_1}\cdots K^{m_s}_{\pi_s}
   G^{(r_0+s)}_{\pi_0m_1\cdots m_s}
   \prod_{j=1}^s\prod_{m=2}^{r_j}\delta_{x_{\pi_j(1)},x_{\pi_j(m)} },
\autoeq$$
where the argument of $G$ has been omitted for brevity, thus
$$G^{(r_0+s)}_{\pi_0m_1\cdots m_s}
=G^{(r_0+s)}_{a_{\pi_0(1)}\cdots a_{\pi_0(r_0)}m_1\cdots
m_s}(x_{\pi_0(1)},\ldots,x_{\pi_0(r_0)},x_{\pi_1(1)},\ldots,x_{\pi_s(1)}).
\autoeq$$
The index structure on
$G$ is sufficient to deduce its arguments.

Functionally differentiating $F_s$ and subsequently setting
the couplings to be independent of position gives
$$\eqalign{
   { \delta F_s\over\delta g^{ a_{N+1} } (x_{N+1} ) }
 & \Vline\sub{ \partial_\mu g^{a_i}=0 }=
   (-1)^{N+s+r_0}\sum_{j=1}^s
   K^{m_1}_{\pi_1}\cdots K^{ m_{j-1} }_{\pi_{j-1}}
   \bigl[\partial_{ a_{N+1} }K^{m_j}_{\pi_j}\bigr]
   K^{ m_{j+1} }_{ \pi_{j+1}}\cdots K^{m_s}_{\pi_s}
\cr
 & \hskip 2cm\times G^{(r_0+s)}_{\pi_0m_1\cdots m_s}
   \delta_{x_{N+1},x_{\pi_j(1)}}
   \prod_{j=1}^s\prod_{m=2}^{r_j}\delta_{x_{\pi_j(1)},x_{\pi_j(m)} }
\cr
 & +(-1)^{N+s+r_0}
   K^{m_1}_{\pi_1}\cdots K^{m_s}_{\pi_s}
   \bigl[\delta_{a_{N+1}}G^{(r_0+s)}_{\pi_0 m_1\ldots m_s}\bigr]
   \prod_{j=1}^s\prod_{m=2}^{r_j}\delta_{x_{\pi_j(1)},x_{\pi_j(m)} }
\cr}
\autoeq$$
\newcount\temp\temp=\count90
where $\delta_{a_{N+1}}:={\delta\over\delta g^{a_{N+1}}(x_{N+1}) }$.
{}From the definition (\the\NKdef) we have
$$\partial_b K^{m}_{a_1\cdots a_j}=K^{m}_{ba_1\cdots a_j}
-K^{c}_{a_1\cdots a_j}{K^m}_{cb}.\autoeq$$
We also observe that, from equation (\the\Zdef),
$$\eqalign{
\delta_{a_{N+1}}G^{(r_0+s)}_{\pi_0 m_1\ldots m_s}&=
-G^{(r_0+s+1)}_{\pi_0a_{N+1}m_1\cdots m_s}\cr
&+\sum_{k=1}^{r_0}{K^c}_{a_{\pi_0(k)} a_{N+1}}
G^{(r_0+s)}_{a_{\pi_0(1)}\cdots a_{\pi_0(k-1)}c
a_{\pi_0(k+1)}\cdots a_{\pi_0(r_0)}
m_1\cdots m_s}\delta_{x_{N+1},x_{\pi_0(k)}}\cr
&+\sum_{k=1}^{s}{K^c}_{m_k a_{N+1}}
G^{(r_0+s)}_{a_{\pi_0(1)}\cdots a_{\pi_0(r_0)}
m_1\cdots m_{k-1} c m_{k+1}\cdots m_s}\delta_{x_{N+1},x_{\pi_k(1)}}
\cr}\autoeq$$\newcount\deltag\deltag=\count90
(in general there are other terms on the right hand side here involving
$A_{ab}$,
from equation (\the\doeq), but these only contribute to the final
result if there are exceptional momenta and they will be
omitted from this analysis).
The $(r_0+s+1)$-point function in (\the\deltag) is a shorthand notation for
$$G^{(r_0+s+1)}_{\pi_0a_{N+1}m_1\cdots m_s}=
G^{(r_0+s+1)}_{a_{\pi_0(1)}\cdots a_{\pi_0(r_0)}a_{N+1}m_1\cdots m_s}
(x_{\pi_0(1)},\ldots,x_{\pi_0(r_0)},x_{N+1},x_{\pi_1(1)},\ldots,x_{\pi_s(1)}).$$

Thus both the terms on the right hand side of equation (\the\temp)
involve
$$K^{m_1}_{\pi_1}\cdots K^{m_s}_{\pi_s}
\sum_{k=1}^s{K^c}_{m_k a_{N+1}}
G^{(r_0+s)}_{\pi_0m_1\cdots m_{k-1} c m_{k+1} m_s},
\autoeq$$
but with opposite sign so that they cancel.

Equation (\the\deltag) can now be re-arranged as
$$\eqalign{
  & { \delta F_s\over\delta g^{ a_{N+1} } (x_{N+1} ) }
  \Vline\sub{ \partial_\mu g^{a_i}=0 }=
   (-1)^{N+1+s+r_0}K^{m_1}_{\pi_1}\cdots K^{m_s}_{\pi_s}
   G^{(r_0+s+1)}_{\pi_0a_{N+1}m_1\cdots m_s}
   \prod_{j=1}^s\prod_{m=2}^{r_j}\delta_{x_{\pi_j(1)},x_{\pi_j(m)} }
\cr
 &+ (-1)^{N+s+r_0}\sum_{j=1}^s
   K^{m_1}_{\pi_1}\cdots K^{ m_{j-1} }_{\pi_{j-1}}
   K^{m_j}_{\pi_ja_{N+1}}
   K^{ m_{j+1} }_{ \pi_{j+1}}\cdots K^{m_s}_{\pi_s}
   G^{(r_0+s)}_{\pi_0m_1\cdots m_s}
\cr
 & \hskip 6cm\times\delta_{x_{N+1},x_{\pi_j(1)}}
   \prod_{j=1}^s\prod_{m=2}^{r_j}\delta_{x_{\pi_j(1)},x_{\pi_j(m)} }
\cr
 & +(-1)^{N+s+r_0}
   K^{m_1}_{\pi_1}\cdots K^{m_s}_{\pi_s}
\sum_{k=1}^{r_0}K^{m_{s+1}}_{a_{\pi_0(k)} a_{N+1}}
G^{(r_0+s)}_{a_{\pi_0(1)}\cdots a_{\pi_0(k-1)}
m_{s+1}a_{\pi_0(k+1)}\cdots a_{\pi_0(r_0)}m_1\cdots m_s}
\cr
 & \hskip 6cm\times\delta_{x_{N+1},x_{\pi_0(k)}}
   \prod_{j=1}^s\prod_{m=2}^{r_j}\delta_{x_{\pi_j(1)},x_{\pi_j(m)} }
\cr}
\autoeq$$
where the summation variable $c$ has been replaced in a suggestive
manner by $m_{s+1}$.

We now note that the indices on $G^{(r_0+s)}$ can be permuted,
provided that one also understands the arguments to be permuted as
well, thus
$$\eqalign{
   &G^ {(r_0+s)} _ { a_{\pi_0(1)}\cdots a_{\pi_0(k-1)}
   m_ {s+1} a_{\pi_0(k+1)}\cdots a_{\pi_0(r_0)}m_1\cdots m_s }
\cr
 &\kern -35pt =G^ {(r_0+s)} _ {a_{\pi_0(1)}\cdots a_{\pi_0(k-1)}
   m_ {s+1} a_{\pi_0(k+1)}\cdots a_{\pi_0(r_0)}m_1\cdots m_s }
   (x_ {\pi_0(1)} ,..,x_ {\pi_0(k-1)} ,x_ {N+1} ,
   x_ {\pi_0(k+1)},..,x_ {\pi_0(r_0)} ,x_ {\pi_1(1)} ,..,x_ {\pi_s(1)})
\cr
 & \kern -35pt =G^ {(r_0+s)} _ {a_{\pi_0(1)}\cdots a_{\pi_0(k-1)}
   a_{\pi_0(k+1)}\cdots a_{\pi_0(r_0)}m_1\cdots m_sm_ {s+1} }
   (x_ {\pi_0(1)} ,..,x_ {\pi_0(k-1)} ,
   x_ {\pi_0(k+1) },..,x_ {\pi_0(r_0) },x_ {\pi_1(1) },
   ..,x_ {\pi_s(1)} ,x_ {N+1} )
\cr
 & =G^ {(r_0+s)} _{ a_{\pi_0(1)}\cdots a_{\pi_0(k-1)}
   a_{\pi_0(k+1)}\cdots a_{\pi_0(r_0)}m_1\cdots m_sm_ {s+1} }.
\cr}\autoeq$$
Thus, introducing an overall minus sign, we arrive at
$$\eqalign{
& -{ \delta F_s\over\delta g^{ a_{N+1} } (x_{N+1} ) }
  \Vline\sub{ \partial_\mu g^{a_i}=0 }=
   (-1)^{N+1+s+r_0+1}K^{m_1}_{\pi_1}\cdots K^{m_s}_{\pi_s}
   G^{(r_0+s+1)}_{\pi_0a_{N+1}m_1\cdots m_s}
   \prod_{j=1}^s\prod_{m=2}^{r_j}\delta_{x_{\pi_j(1)},x_{\pi_j(m)} }
  \cr
 &+ (-1)^{N+1+s+r_0}\sum_{j=1}^s
   K^{m_1}_{\pi_1}\cdots K^{ m_{j-1} }_{\pi_{j-1}}
   K^{m_j}_{\pi_ja_{N+1}}
   K^{ m_{j+1} }_{ \pi_{j+1}}\cdots K^{m_s}_{\pi_s}
   G^{(r_0+s)}_{\pi_0m_1\cdots m_s}
\cr
 & \hskip 6cm \times \delta_{x_{N+1},x_{\pi_j(1)}}
   \prod_{j=1}^s\prod_{m=2}^{r_j}\delta_{x_{\pi_j(1)},x_{\pi_j(m)} }
\cr
 & +(-1)^{N+1+s+1+r_0+1}
   K^{m_1}_{\pi_1}\cdots K^{m_s}_{\pi_s}
\sum_{k=1}^{r_0}K^{m_{s+1}}_{a_{\pi_0(k)} a_{N+1}}
G^{(r_0+s)}_{a_{\pi_0(1)}\cdots a_{\pi_0(k-1)}
a_{\pi_0(k+1)}\cdots a_{\pi_0(r_0)}m_1\cdots m_sm_{s+1}}
\cr
 & \hskip 6cm\times\delta_{x_{N+1},x_{\pi_0(k)}}
   \prod_{j=1}^s\prod_{m=2}^{r_j}\delta_{x_{\pi_j(1)},x_{\pi_j(m)} }.
\cr}
\autoeq$$

Using this in equation (\the\Npointx) and (\the\Npopoint)
and re-arranging the summations one sees the desired
structure emerging, but there are more terms to be taken into
account. These come from extra contributions to the
regularised Green functions (\the\Npointx) when
$\partial_\mu g^b\ne 0$. One only need consider the terms linear in
$\partial_\mu g^b\ne 0$, as higher order contributions vanish
when the condition $\partial_\mu g^b=0$ is imposed after one functional
differentiation. These extra terms have the effect of symmetrising the
result between all $N+1$ indices, and the full expression (\the\Npointx)
is recovered with $N$ replaced by $N+1$.
The $A, B$ and $C$ terms can be verified without
diffuclty.

The form of equation (\the\Npointp) is basically the same in dimensions
other than four and/or when masses are included - all that changes
are the terms involving $A,B$ and $C$.
\vskip 1cm
{\bf Appendix 2}
\vskip .5cm
In this appendix we give the full expression for the
co-variant renormalisation
group equation for regularised four point functions, not just at the
symmetric point. The
derivation is a straightforward, but tedious, application of the
techniques described in the text. The result, in momentum space, is
$$\eqalign{
 &\left[
        \Bigl(\kappa{\partial\over\kappa}\Vline\sub{g} + {\cal L}_\beta\Bigr)
        G^R(p,q,r,s)
  \right]_{abcd}
 =\left[ {\tau_{ab}}^f
         G^R_{fcd}(p+q,r,s) + \; 5\;
         \hbox{terms}
  \right]
  \cr
     &-\left[ \bigl(\Del_a{\tau_{bc}}^f\bigr)
                G^R_{fd}(-s,s)+
            \left\{
              \matrix{b\leftrightarrow d\cr
                      q\leftrightarrow s\cr}
             \right\}
            +\left\{
              \matrix{c\leftrightarrow d\cr
                      r\leftrightarrow s\cr}
             \right\}\right]
           -\bigl(\Del_b{\tau_{cd}}^f\bigr)G^R_{fa}(-p,p)
  \cr
     &     -{\kappa^{-\epsilon}\over 2}
           \Bigl(p^2q^2\Del_a\Del_b\chi_{cd}+ \; 11\;
           \hbox{terms}\Bigr)
\cr
     &     -{\kappa^{-\epsilon}\over 2}\left[\Bigl(s^2(q^2+p^2+4q.p){R^e}_{bca}
           +s^2(r^2+p^2+4r.p){R^e}_{cba}\Bigr)\chi_{ed}+
           \left\{
            \matrix{b\leftrightarrow d\cr
                    p\leftrightarrow s\cr}
           \right\}+
           \left\{
            \matrix{c\leftrightarrow d\cr
                    r\leftrightarrow s\cr}
           \right\}\right]
  \cr
     &-{\kappa^{-\epsilon}\over 2}\Bigl(p^2(q^2+r^2+4q.r){R^e}_{cdb}
     +p^2(q^2+s^2+4q.s){R^e}_{dcb}\Bigr)\chi_{ea}
  \cr
     &-\kappa^{-\epsilon}\bigl[
     (q.r)s^2\Del_a{\chi_{bcd}} + 11\;\hbox{permutations}\bigr]
     -\kappa^{-\epsilon}\bigl[2(p.q)(r.s)\chi_{abcd}
     + 2\;\hbox{permutations}\bigr].
  \cr
}\autoeq$$\newcount\Atwo\Atwo=\count90

Once again the structure of the $\chi$-terms is peculiar to massless
theories in four dimensions, but the $\tau$-terms are the same
for all theories.

\vfill\eject
{\bf References}\hfill
\vskip .5cm
\item{[\the\KZ]} V.G. Knizhnik and A.B. Zamolodchikov,
Nucl. Phys. {\bf B247}, (1984), 83
\smallskip
\item{[\the\Kutasov]} D. Kutasov, Phys. Lett. {\bf 220B}, (1989), 153
\smallskip
\item{[\the\Sonoda]} H. Sonoda, Nucl. Phys. {\bf B383}, (1992), 173;
{\bf B394}, (1993), 302
\smallskip
\item{[\the\Zwiebach]} K. Ranganathan, Nucl. Phys. {\bf 408B}, (1993),
180\hfil\break
K. Ranganathan, H. Sonoda and B. Zwiebach, MIT-CTP-2193, April 1993
\smallskip
\item{[\the\Denjoe]} {\sl Geometry, The Renormalisation Group and Gravity}
D. O'Connor and C.R. Stephens\hfil\break
in \sl Directions In General Relativity, \rm Ed. B.L. Hu, M.P. Ryan Jr. and
C.V. Vishveshwava\hfil\break
Proceedings Of The 1993 International Symposium, Maryland, Vol. 1, C.U.P.
(1993)
\smallskip
\item{[\the\Ross]} D. Ross and C.A. L\"utken, Phys. Rev. {\bf B45}, (1990),
11837
\smallskip
\item{[\the\ZamA]} A.B. Zamolodchikov,
Pis'ma Zh. Eksp. Teor. Fiz. {\bf 43}, (1986), 565
\smallskip
\item{[\the\Lassig]} M. L\"assig, Nucl. Phys. {\bf B334}, (1990), 652
\smallskip
\item{[\the\Dolan]} B.P. Dolan, To appear in Int. J. of Mod. Phys. A
\smallskip
\item{[\the\HughIan]} I. Jack and H. Osborn, Nucl. Phys. {\bf B343}, (1990),
647
\smallskip
\item{[\the\ZamB]} A.B. Zamolodchikov, Rev. Math. Phys. {\bf 1}, (1990), 197
\smallskip
\item{[\the\NLSR]} G.M. Shore, Nucl. Phys. {\bf B362}, (1991), 85
\smallskip
\item{[\the\Kobayashi]} {\sl Foundations of Differential Geometry:  Vol.I}
S. Kobayashi and K. Nomizu\hfil\break
1969 Wiley (New York)
\smallskip
\item{[\the\Brown]} L.S.~Brown, Ann. Phys. \bf 126\rm, (1980), 135\hfil\break
L.S.~Brown and J.C.~Collins, Ann. Phys. \bf 130\rm, (1980), 215

\bye